\documentclass[
    ,final            
  ]
  {aipproc}

\layoutstyle{6x9}

\begin{document}

\title{
{\noindent \small UNITU-THEP-16/2004
\hspace*{1cm} IPPP/04/75  
\hspace*{1cm} DCPT/04/150 
\hfill hep-ph/0411347}\\
{~}\\
Dynamical Chiral Symmetry Breaking in Landau gauge QCD}

\classification{12.38.Aw 14.65.Bt 14.70.Dj 12.38.Lg 11.30.Rd 11.15.Tk 02.30.Rz}
\keywords      {Confinement, dynamical chiral symmetry breaking, 
gluon propagator, quark propagator, Dyson-Schwinger equations}

\author{C.~S.~Fischer}{
  address={IPPP, University of Durham, Durham DH1 3LE, U.K.}
}

\author{R.~Alkofer}{
  address={Institute for Theoretical Physics, University of
       T\"ubingen, D-72076 T\"ubingen, Germany}
}

\begin{abstract}
We summarise results for the propagators of Landau gauge QCD from 
the Green's functions approach and lattice calculations. The 
nonperturbative solutions for the ghost, gluon and quark propagators 
from a coupled set of Dyson-Schwinger equations agree almost 
quantitatively with corresponding lattice results. Similar unquenching 
effects are found in both approaches. The dynamically generated quark 
masses are close to `phenomenological' values. The chiral condensate
is found to be large. 
\end{abstract}

\maketitle



The infrared behaviour of the propagators of Landau gauge QCD has been investigated
extensively over the past years in
lattice Monte Carlo simulations and the
continuum Green's functions approach. Lattice simulations are the only ab initio 
method known so far and are by now precise enough to pin down these propagators 
accurately in a large momentum range centered around 1 GeV. In the deep infrared,
however, lattice results are inevitably plagued by finite volume effects. In the 
continuum formulation of QCD the Dyson-Schwinger equations (DSEs) provide a tool 
complementary to lattice simulations. They can be solved analytically in the 
infrared. Furthermore numerical solutions over the whole momentum range are 
available by now. The truncation assumptions necessary to close the DSEs can be
checked in the momentum regions where lattice results are available. In general,
results from DSEs have the potential to provide a sucessful description of hadrons 
in terms of quarks and gluons, see \cite{Maris:2003vk,Alkofer:2000wg,Roberts:2000aa} 
and references therein. 

The ghost, gluon and quark propagators, $D_G(p)$, $D_{\mu \nu}(p)$ and $S(p)$, 
in Euclidean momentum space can be generically written as
\begin{eqnarray}
  D_G(p) &=& - \frac{G(p^2)}{p^2} \,,
  \label{ghost_prop}\\
  D_{\mu \nu}(p) &=& \left(\delta_{\mu \nu} - \frac{p_\mu
      p_\nu}{p^2} \right) \frac{Z(p^2)}{p^2} \, ,
  \label{gluon_prop} \\
  S(p) &=& \frac{1}{-i  p\!\!\!/\, A(p^2) + B(p^2)}
  =  \frac{Z_Q(p^2)}{-ip\hspace{-.5em}/\hspace{.15em}+M(p^2)}
  \, .
  \label{quark_prop}
\end{eqnarray}
Here we have chosen Landau gauge which is a fixed point under renormalization
\cite{Ellwanger:1995qf}. The Dyson-Schwinger equations for the ghost and gluon 
dressing functions, $G(p^2)$ and $Z(p^2)$, have been investigated in refs.\
\cite{vonSmekal:1997is,Fischer:2002hn}. They 
can be solved analytically in the infrared and one finds simple power laws,
\begin{eqnarray}
  Z(p^2) &\sim& (p^2)^{2\kappa}, \nonumber\\
  G(p^2) &\sim& (p^2)^{-\kappa},
  \label{g-power}
\end{eqnarray}
for the gluon and ghost dressing function with exponents related to
each other. The relations 
(\ref{g-power}) can be determined from the ghost-DSE alone
and are independent of the truncation scheme. The exponent $\kappa$ is an 
irrational number and depends only slightly on the dressing of the ghost-gluon
vertex \cite{Lerche:2002ep,Zwanziger:2001kw}. With a bare vertex one obtains
$\kappa = (93 - \sqrt{1201})/98 \approx 0.595$. Recently these results 
have been confirmed independently in studies
of the exact renormalisation group equation \cite{Pawlowski:2003hq}.

\begin{figure}[t!]
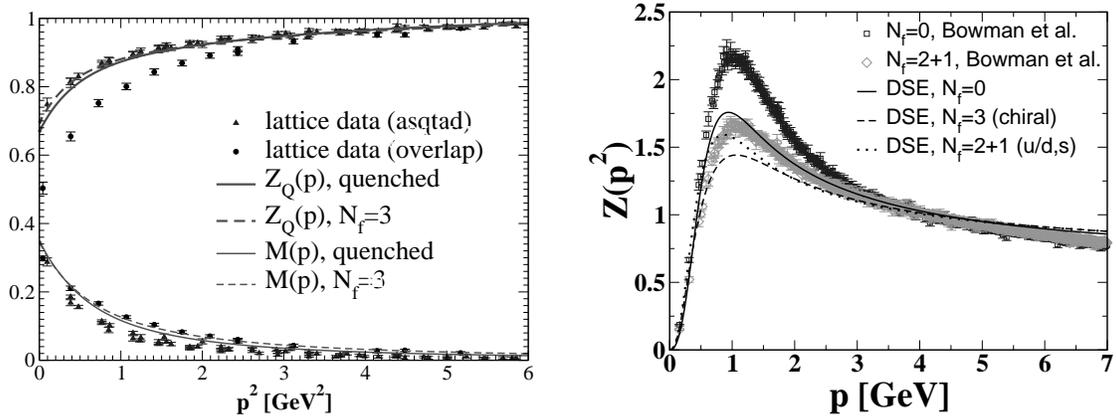

\includegraphics[width=7cm]{quark_ZM_lin.adel04.eps}
\hspace*{0.5cm}
\includegraphics[width=7cm]{lattice_glue_unq4.eps}
\caption{Left: The quenched and unquenched quark mass function $M(p^2)$ and the 
wave function $Z_Q(p^2)$ from the DSE approach \cite{Fischer:2003rp}
compared to results from quenched lattice calculations \cite{LATTICE_QUARK}.
Right: The quenched and unquenched gluon dressing function from the DSE 
approach \cite{Fischer:2003rp} compared to results from unquenched lattice 
calculations \cite{Bowman:2004jm}.}
\label{fig}
\end{figure}

The dynamical generation of quark masses can be studied in the Dyson-Schwinger 
equation for the quark propagator. It is a genuinely non-perturbative phenomenon 
and requires a careful treatment of the quark-gluon interaction. In ref. 
\cite{Fischer:2003rp} we demonstrated that sizeable nontrivial Dirac-structures 
in the quark-gluon vertex are necessary to generate dynamical
quark masses of the order of 300-400 MeV. Our results for the quenched quark mass 
function $M(p^2)$ and the wave function
$Z_Q(p^2)$ are compared to the quenched lattice results of refs.
\cite{LATTICE_QUARK} in fig.~\ref{fig}. The overall qualitative and quantitative 
agreement between both approaches is very good. The DSE results are within the 
bounds given by the two different formulations of fermions on the lattice. 

Including the backreaction of the quark-propagator on the ghost and gluon system
leads to a coupled set of three Dyson-Schwinger equations for the propagators of 
QCD. These equations have been solved in \cite{Fischer:2003rp} and allowed a 
prediction of possible effects of unquenching QCD on the propagators. As can be seen
from fig. \ref{fig} including $N_f=3$ chiral quarks in the gluon
DSE hardly changes the results for the quark propagator. 
The chiral condensate is nearly unaffected. It will be 
interesting to compare these results to unquenched lattice calculations when
available.

Unquenched lattice results for the gluon propagator including the effects of 
two light (up-) and one heavy (strange-) quark have been published recently
\cite{Bowman:2004jm} and are compared to the corresponding results from our
DSE-approach in fig. \ref{fig}. The screening effect from the quark loop is
clearly visible in the lattice results for momenta $p$ larger than $p=0.5$ GeV: 
the gluonic self interaction becomes less important in this region and the gluon
dressing increases. This effect can also be seen in the DSE-approach. In the quenched
case there is a discrepancy between the DSE-result and the lattice data, which 
can be traced back to the fact that not all effects from the gluonic self 
interaction are accounted for in the DSE truncation. When this part 
of the gluon interaction becomes less dominant
in the unquenched case, both the lattice and the DSE-approach agree very well on
a quantitative level, provided similar bare quark masses are taken into account.
In the chiral limit the screening effect of the quark loop becomes even stronger
as can be seen from the DSE-results in fig.~\ref{fig}. This is expected 
as the energy needed to create a quark pair out of the vaccuum becomes smaller with
decreasing bare quark mass. 

Both, the lattice calculations and the Green's functions approach agree in the fact 
that unquenching does not affect the extreme infrared of the ghost and gluon 
propagators. Again, this is easily explained from dynamical chiral 
symmetry breaking: there is
not enough energy to generate a quark pair from the vacuum below a certain threshold.
Then the quark degrees of freedom decouple from the Yang-Mills sector of the theory.


\begin{theacknowledgments}
We thank the organizers of {\it Quark Confinement and the Hadron Spectrum VI\/}
for their efforts which made this very inspiring conference possible.
We are grateful to D.~Leinweber, 
F.~Llanes-Estrada, M.~Pennington, P.~Tandy and A.~Williams
for helpful discussions.  
This work has been supported by the Deutsche For\-schungsgemeinschaft
(DFG) under contract Fi 970/2-1.
\end{theacknowledgments}

\end{document}